\documentclass[twocolumn,times]{aastex63}
\turnoffeditone
\usepackage{graphicx,natbib,amsmath,gensymb,multirow}
\bibpunct{(}{)}{;}{a}{}{,}

\newcommand{\narrowfig}[3]{%
\begin{figure}[tbp]
\begin{center}
\includegraphics[width=86mm]{#1}
\caption{#3}
\label{#2}
\end{center}
\end{figure}
}

\newcommand{\widefig}[3]{%
\begin{figure*}[tbp]
\begin{center}
\includegraphics[width=180mm]{#1}
\caption{#3}
\label{#2}
\end{center}
\end{figure*}
}

\begin{document}

\title{Design and Performance Analysis of a Highly Efficient Polychromatic Full-Stokes Polarization Modulator for the CRISP Imaging Spectrometer}
\shorttitle{The CRISP Polychromatic Modulator}

\author[0000-0002-5084-4661]{A.G.~de~Wijn}
\affil{High Altitude Observatory, National Center for Atmospheric Research, P.O. Box 3000, Boulder, CO 80307, USA}

\author[0000-0002-4640-5658]{J.~de~la~Cruz~Rodr\'\i{}guez}
\author[0000-0002-2281-8140]{G.B.~Scharmer}
\author{G.~Sliepen}
\author{P.~S\"utterlin}
\affil{Institute for Solar Physics, Department of Astronomy, Stockholm University, AlbaNova University Centre, SE-106 91, Stockholm,
Sweden}

\correspondingauthor{A.G.~de~Wijn}
\email{dwijn@ucar.edu}

\shortauthors{De Wijn et al.}

\begin{abstract}
	We present the design and performance of a polychromatic polarization modulator for the CRisp Imaging SpectroPolarimeter (CRISP) Fabry-Perot tunable narrow-band imaging spectropolarimer at the Swedish 1-m Solar Telescope (SST).
	We discuss the design process in depth, compare two possible modulator designs through a tolerance analysis, and investigate thermal sensitivity of the selected design.
	\edit1{The trade-offs and procedures described in this paper are generally applicable in the development of broadband polarization modulators.}
	The modulator was built and has been operational since 2015.
	Its measured performance is close to optimal between 500 and 900~nm, and differences between the design and as-built modulator are largely understood.
	We show some example data, and briefly review scientific work that used data from SST/CRISP and this modulator.
\end{abstract}

\keywords{instrumentation: polarimeters}

\section{Introduction}\label{sec:introduction}
Our knowledge of solar magnetism relies heavily on our ability to detect and interpret the polarization signatures of magnetic fields in solar spectral lines.
Consequently, new Stokes polarimeters are designed to have the capability to observe the solar atmosphere in a variety of spectral lines over a wide wavelength range.
One immediate instrument requirement stemming from this need for wavelength diversity is that the polarization modulation scheme must be \emph{efficient} at all wavelengths within the working range of the spectropolarimeter.
(For the definition of polarimetric efficiency see \citealp{2000ApOpt..39.1637D}.)
Typically, one attempts to achieve this goal by achromatizing the polarimetric response of a modulator.
This, for instance, is the rational behind the design of super-achromatic wave plates \citep{1974MExP...12..361S,2004JQSRT..88..319S,2008ChJAA...8..349M}.
\cite{2010ApOpt..49.3580T} argued that for many instruments achromaticity is too strong a constraint, and instead proposed the concept of the \emph{polychromatic} modulator that is efficient at all wavelengths of interest, but has polarimetric properties that vary with wavelength.

In this paper, we present the development process of a modulator for the CRisp Imaging SpectroPolarimeter (CRISP) Fabry-Perot tunable narrow-band imaging instrument \citep{2006A&A...447.1111S,2008ApJ...689L..69S} at the Swedish 1-m Solar Telescope \citep[SST,][]{2003SPIE.4853..341S}.
First, we compare the performance of two possible modulator designs, and use a Monte-Carlo tolerance analysis to evaluate their robustness.
We analyze the sensitivity of the modulator to thermal conditions, and present the opto-mechanical packaging and electrical interfaces.
The modulator was constructed and tested at the High Altitude Observatory (HAO).
We compare as-built properties to those of the design.
Finally, we show some example polarimetric observations made using this modulator.

This modulator was designed and built to replace a modulator based on Liquid Crystal Variable Retarders (LCVRs).
LCVRs are electro-optical devices that have a fixed fast axis orientation, but, as the name implies, can be set to any retardance within some range by applying an AC voltage.
LCVRs generally have much slower switching speeds than Ferro-electric Liquid Crystals (\edit1{FLCs}).
In contrast to LCVRs, \edit1{FLCs} have a constant retardance but switch their fast axis orientation between two states separated by a switching angle, typically around $45\degr$.
The LCVRs in the old CRISP modulator had to be ``overdriven'' and the modulator state order had to be optimized in order to switch during the $10~\mathrm{ms}$ readout time of the CRISP cameras.
More importantly, however, thermal sensitivities of the setup forced polarimetric calibration more frequently than desired \citep{2008A&A...489..429V}.

We limit ourselves to designs that use \edit1{FLCs} because their fast switching speed allows the state of the modulator to be changed in less than the allotted $10~\mathrm{ms}$, thus allowing for the highest possible modulation rate.
Fast modulation is desirable because seeing-induced crosstalk between Stokes parameters that is a dominant source of error in ground-based polarimeters is less at higher modulation rates \citep{1987ApOpt..26.3838L,2004ApOpt..43.3817J,2012ApJ...757...45C}.
Also, it is of importance in maximizing the overall efficiency of the polarimeter.
Many present-day CCD and CMOS detectors allow simultaneous exposure and readout \edit1{so} that the switching speed of the modulator becomes the limitation in the overall duty cycle---and thus efficiency---of the modulator.

\edit1{For a recent review of instrumentation for solar spectropolarimetery we direct the reader to \cite{2019OptEn..58h2417I}.
\cite{2014SPIE.9099E..0LR} present a more general review of instrumentation for measurements of polarized light.}

\section{Design}\label{sec:design}

A computer program was developed at HAO to determine component parameters for a given modulator design \citep{2010ApOpt..49.3580T}.
This program was used successfully to design the modulators for the ProMag, CoMP-S, SCD, ChroMag, and UCOMP instruments built or under construction by HAO \citep{2008SPIE.7014E..16E, Kucera2010, 2015IAUGA..2246687K, 2012SPIE.8446E..78D}.
More recently, it, or similar programs derived from it or independently implemented by others, have been used to design modulators, e.g., for the DKIST \citep{2018JATIS...4d4006H}.
The code can use several different merit functions.
We choose to minimize the maximum of the deviation of the modulation efficiency \edit1{$\epsilon_Q$, $\epsilon_U$, and $\epsilon_V$} in Stokes $Q$, $U$, and $V$ from the optimal value of $1/\sqrt{3}$ for balanced modulation at a number of user-specified wavelengths, normalized by the efficiency \edit1{$\epsilon_I$} in Stokes $I$.
\edit1{It is possible to bias the modulation efficiency to prefer linear or circular polarization.
However, such schemes are of limited use because the SST is not a polarization-free telescope \citep{Selbing2005}.}

We study two designs: one consisting of two \edit1{FLC} devices followed by one fixed retarder that we will refer to as the FFR design, and one consisting of an \edit1{FLC}, a fixed retarder, a second \edit1{FLC}, and a second fixed retarder that we will refer to as the FRFR design.
The HAO-designed instruments mentioned above all use the FFR design.
Others have implemented FRFR designs \citep[e.g.,][]{1999OptEn..38.1402G, 2001ASPC..236...16K,2016A&A...590A..89I}
The FFR design has 5 free parameters, whereas the FRFR design has 7 (see Table~\ref{tab:designs}).
Both have significant freedom to optimize the design over wide wavelength ranges.

All modulators discussed here were optimized for balanced modulation, i.e., equal efficiency in $Q$, $U$, and $V$, at 16~equidistant wavelengths over the 500---900~nm operating wavelength range of the CRISP instrument.
We allow the program to choose the retardances of the \edit1{FLCs} and the retarders, as well as the orientations of the second \edit1{FLC} and the retarders.
Experience has shown that the best configurations have an orientation very close to $0$ or $90\degr$ with respect to the orientation of the analyzing polarizer for \edit1{the bisector of} the first \edit1{FLC}.
We therefore fix the orientation of the first \edit1{FLC} at 0~degrees to eliminate one free parameter.
The switching angle of an \edit1{FLC} is sensitive to both temperature and drive voltage \citep{Gisler2005,2003SPIE.4843...45G}.
Hence, we assume that we can \edit1{adjust the drive voltage at a given operating temperature so that} the \edit1{FLC} switching angles \edit1{are} 45~degrees.
We also account for dispersion of birefingence for all elements of the modulator \edit1{using measurements from similar optics}.

The program can use several different optimization techniques.
We first use a Latin Hypercube Sampling algorithm \citep{10.2307/1268522} to probe the parameter space.
We use a large population size of 25,000 but only 5 iterations in which we shrink the parameter space around the best solution.
We then apply a downhill simplex method \citep{10.1093/comjnl/7.4.308} to refine the solution.
To increase confidence that we did not find a local minimum, we repeat the search several times and check that we consistently find the same solution.
The resulting designs are summarized in Table~\ref{tab:designs}.

\begin{table}[tbp]
	\caption{\label{tab:designs}FRFR and FFR modulator designs.
	The orientation value of the \edit1{FLCs} refers to the bisector of the two fast axis positions that are separated by $45\degr$.}
\centering
\begin{tabular}{lrr}
\hline
Component  & Retardance & Orientation \\
           & waves at 665~nm & degrees\\ \hline\hline
\multicolumn{3}{c}{FRFR}              \\ \hline
\edit1{FLC} 1     & $0.429$    & $0$         \\
Retarder 1 & $0.181$    & $121.0$     \\
\edit1{FLC} 2     & $0.324$    & $17.5$      \\
Retarder 2 & $0.543$    & $109.6$     \\ \hline\hline
\multicolumn{3}{c}{FFR}               \\ \hline
\edit1{FLC} 1     & $0.490$      & $0$       \\
\edit1{FLC} 2     & $0.248$      & $112.9$   \\
Retarder 1 & $0.228$      & $108.8$   \\
\end{tabular}
\end{table}

The next step is to evaluate the robustness of the design using a Monte-Carlo tolerancing method.
Efficiencies were calculated for a total of 1000 modulator realizations with parameters chosen from a uniform distribution around the design values.
The width of the distributions was chosen to be the vendor-supplied accuracies for the retardances of the devices of $50~\mathrm{nm}$ for the \edit1{FLCs} and $2~\mathrm{nm}$ for the retarder.
\edit1{For the orientation we assume an error of up to 1~degree.
Experience has shown that alignment of the optics with this accuracy is possible by hand with a simple lab setup.}
The switching angle of the \edit1{FLCs} is assumed to be 45~degrees with insignificant error.
\edit1{FLCs} typically have large manufacturing errors in their retardances.
Since as-built retardances will be known prior to assembly of the modulator, the tolerancing process re-optimizes the angles of the components after the retardances have been chosen.

As an example, a realization of the FFR modulator might have \edit1{FLCs} with retardances of $0.513$ and $0.236$ waves, and a retarder with a value of $0.243$ waves, at the reference wavelength of $665~\mathrm{nm}$.
The tolerancing procedure would then re-optimize the modulator design to find optimal angles of $115.5$ and $109.9\degr$ for the second \edit1{FLC} and the retarder.
Then, the procedure will perturb all the angles to account for mounting errors, to, say, $-0.6$, $114.8$, and $109.8\degr$, and finally calculate the efficiencies of this modulator realization.

The resulting expected modulator performance is shown in Figs.~\ref{fig:FRFR} and~\ref{fig:FFR}.
An even better result can be achieved by re-optimizing the retardances of the fixed retarders in addition to the orientations after the as-built \edit1{FLC} retardances are known.
This was not pursued for the CRISP modulator due to time constraints, and because the design is shown to be tolerant to expected manufacturing errors.

Figures~\ref{fig:FRFR} and~\ref{fig:FFR} show that both designs are well-behaved.
The nominal FRFR design exhibits better overall performance than the FFR design, which is not surprising in view of its higher number of degrees of freedom.
The tolerance analysis shows that the FRFR design is considerably less resistant to manufacturing errors than the FFR design, particularly in $\epsilon_V$ between 500 and 800~nm.
We show this design here to demonstrate the importance of performing a tolerance analysis.
It is possible to find other FRFR designs that have slightly worse performance, but are more robust against manufacturing errors.
However, the FFR design performs very well over this wavelength range and has the benefit of one less component, and thus results in a thinner stack of optics with fewer interfaces.
In our case, we select the FFR design primarily because the modulator must fit in a tight space in the existing CRISP optical setup.

\narrowfig{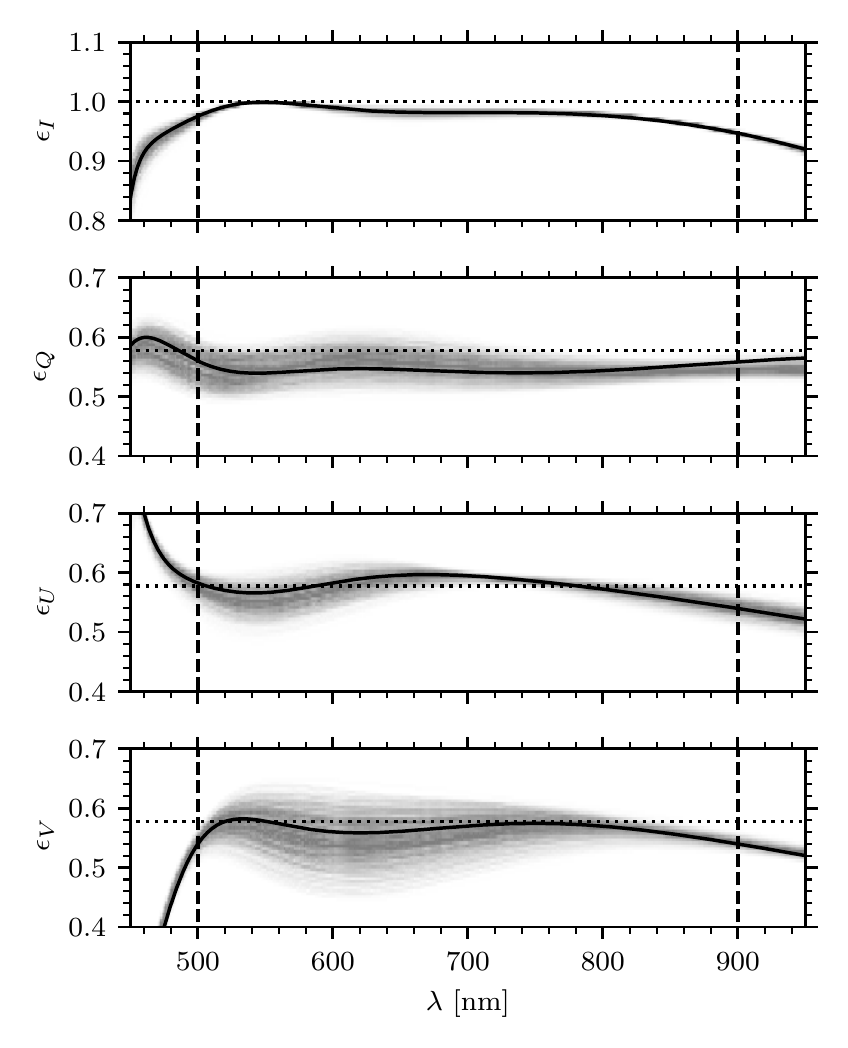}{fig:FRFR}{Theoretical $I$, $Q$, $U$, and $V$ efficiencies with tolerances for the FRFR design.
Solid curves: design performance of the modulator as a function of wavelength.
Horizontal dotted lines: theoretical efficiencies for a perfectly balanced and optimally efficient modulator.
Vertical dashed lines: lower and upper bound of the design wavelength range.
The grayscale background shows the expected spread of performance as a result of component and construction tolerances.}

\narrowfig{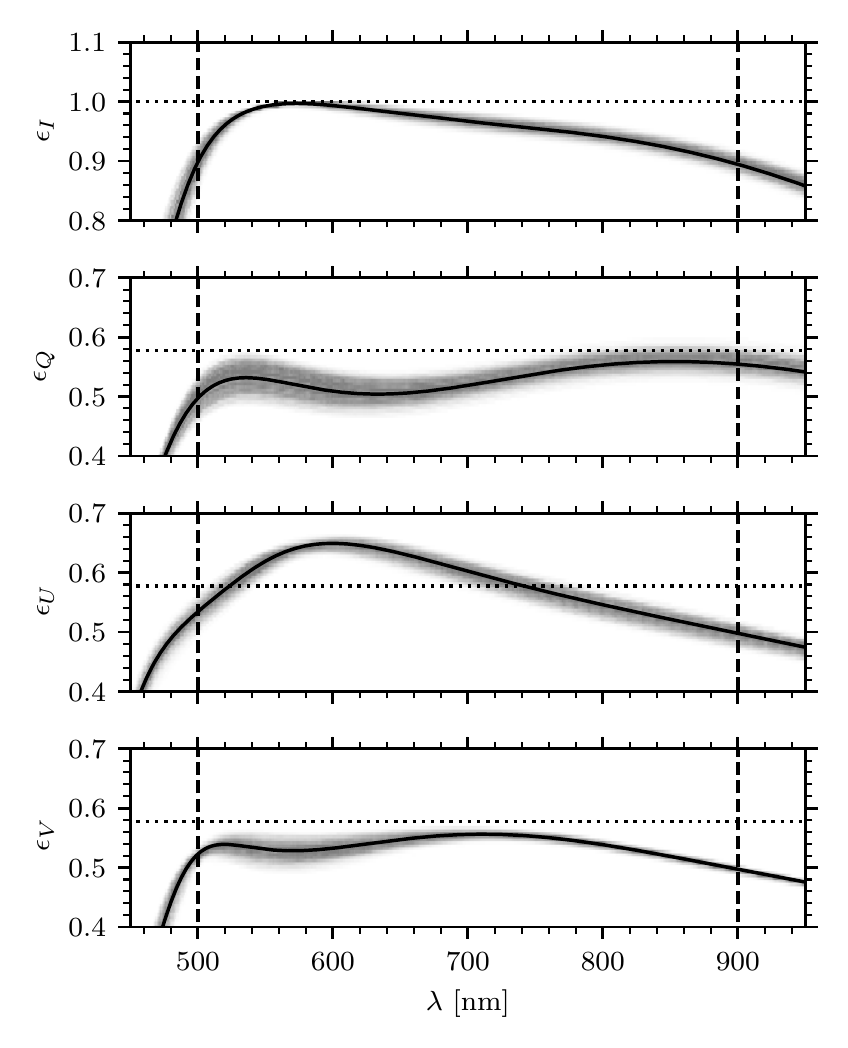}{fig:FFR}{Theoretical $I$, $Q$, $U$, and $V$ efficiencies with tolerances for the FFR design in the same format as Fig.~\ref{fig:FRFR}}

There is considerable freedom to pick a reference wavelength.
Our experience has shown that a wavelength at or slightly below the middle of the operational range is a good choice for practical reasons.
Here, we picked $665~\mathrm{nm}$, also because the program chooses to use \edit1{FLCs} with retardances that are equal to $\lambda/2$ and $\lambda/4$ at that wavelength within the margin of error.
We fix these components at those values \edit1{out of convenience} and optimize the fixed retarder and component orientations.
We find that the orientation of the 2nd \edit1{FLC} does not change.
The fixed retarder value and orientation change slightly to $0.225\lambda$ and $109.1~\mathrm{degrees}$.

\edit1{FLCs with these specifications were procured from Citizen Finetech Miyota.
	The FLC used in these devices is MX8068.
A polycarbonate retarder was procured from Meadowlark Optics.}

\section{Thermal Analysis}

The switching angle of \edit1{FLCs} is somewhat sensitive to temperature.
\cite{2003SPIE.4843...45G} measured it as a function of temperature and found a mostly linear relationship with a coefficient of $-0.4\degr/\mathrm{K}$.
\edit1{This coefficient is specific to the FLC.
\cite{Gisler2005} shows an example measurement with a coefficient of $-0.41\degr/\mathrm{K}$.
For this analysis we use the larger, more conservative value.}

We evaluate the effect of temperature change of the modulator following a procedure similar to \cite{2013SoPh..283..601L}.
\edit1{We follow the notation of \cite{2000ApOpt..39.1637D} and refer the reader to that work for a rigorous mathematical treatment of polarimetric measurements.}
The Stokes vector $\mathbf{S}$ is modulated into a vector of intensities $\mathbf{I}$.
The modulation can be described by a modulation matrix $\mathbf{O}$,
\begin{equation}
	\mathbf{I} = \mathbf{O}\,\mathbf{S}.
\end{equation}
A demodulation matrix $\mathbf{D}$ is used to recover the Stokes vector,
\begin{equation}
	\mathbf{S} = \mathbf{D}\,\mathbf{I}.
\end{equation}
A difference in temperature of the modulator during observations and calibrations will result in a mismatch of the modulation and demodulation matrices.
We denote with \edit1{$\mathbf{O}'$ and $\mathbf{D}'$ the modulation and demodulation matrices} derived from the calibration, and with $\mathbf{S}'$ the inferred Stokes vector,
\begin{equation}
	\mathbf{S}' = \mathbf{D}'\,\mathbf{I}.
\end{equation}
We can then relate the inferred Stokes vector $\mathbf{S}'$ and the real Stokes vector $\mathbf{S}$ through an error matrix,
\begin{equation}
	\mathbf{S} = \mathbf{X}\,\mathbf{S}'.
\end{equation}
It is easy to see that we now have
\begin{equation}
	\mathbf{X}\,\mathbf{D}' = \mathbf{D},
\end{equation}
and using $\mathbf{D}'\,\mathbf{O}' = \mathbf{I}_4$ we find
\begin{equation}
	\mathbf{X} = \mathbf{D}\,\mathbf{O}'.
\end{equation}
\edit1{For our error analysis,} we can calculate the modulation matrix $\mathbf{O}'$ from the unperturbed design, and demodulation matrices $\mathbf{D}$ for several switching angles to determine the permissible change in temperature.

Limits must be imposed on every element of the matrix $\mathbf{X}$.
The diagonal elements represent a scale error that is much less sensitive than crosstalk errors.
The scaling on $I$ is unconstrained after normalizing $\mathbf{S}$ by $I$.
Furthermore, the elements in the $Q$ and $U$ columns can be scaled by the maximum expected linear polarization signal, and those in the $V$ column can be scaled by the maximum expected circular polarization signal.
We follow \cite{2008SoPh..249..233I} and adopt maxima of $e=0.001$ \edit1{of $I$} for the crosstalk error \edit1{between Stokes $Q$, $U$, and $V$}, $a=0.05$ for scale error, $p_\mathrm{l}=15\%$ \edit1{of $I$} for linear polarization, and $p_\mathrm{v}=20\%$ \edit1{of $I$} for circular polarization.
We then find
\begin{equation}\label{eq:xtol}
	|\mathbf{X}-\mathbf{I}_4|\le\begin{pmatrix}
	     &0.333&0.333&0.250\\
	0.001&0.050&0.007&0.005\\
	0.001&0.007&0.050&0.005\\
	0.001&0.007&0.007&0.050
\end{pmatrix},
\end{equation}
\edit1{where, e.g., the second and third element of the top row are given by the ratio $a/p_\mathrm{l}$, and the second and third element of the last column are given by the ratio $e/p_\mathrm{v}$.}

\narrowfig{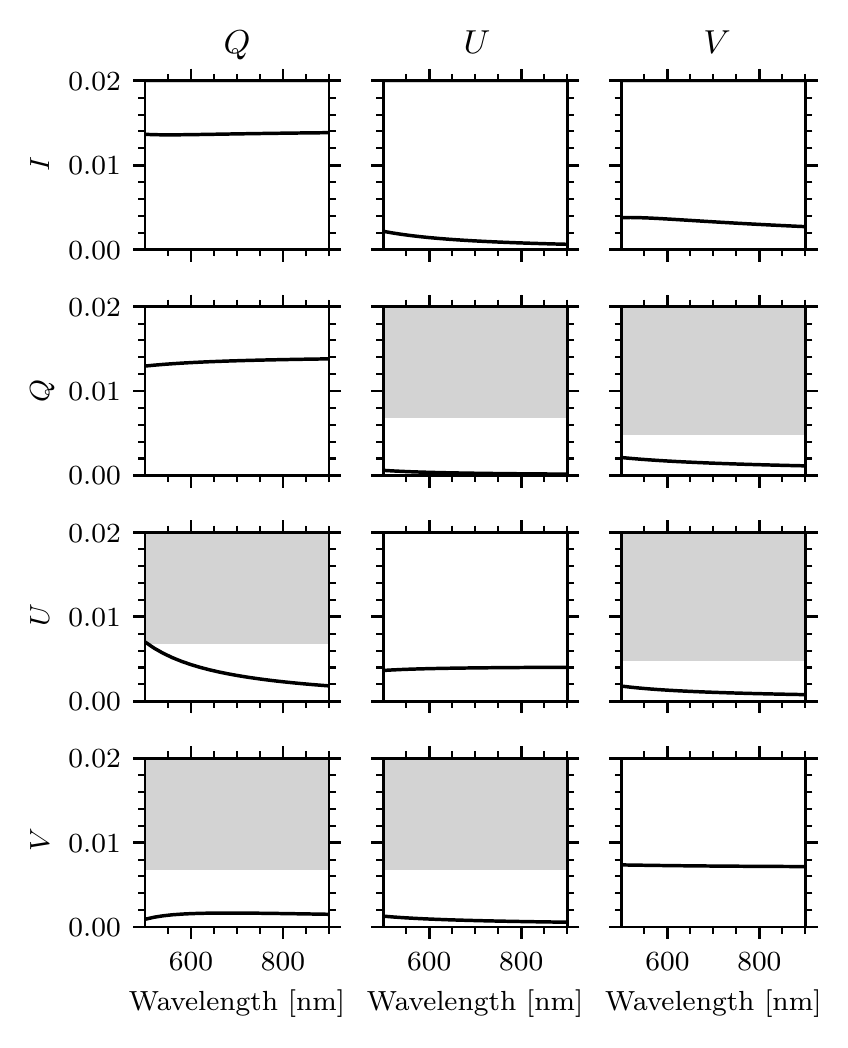}{fig:xmatrix}{The $|\mathbf{X}-\mathbf{I}_4|$ matrix elements for the error introduced by $0.5~\mathrm{K}$ change in temperature assuming a $-0.4\degr/\mathrm{K}$ coefficient for the switching angle of the \edit1{FLCs}.
The first column of the matrix is omitted because it is identically $0$.
Gray areas are outside the limits given in Eq.~\ref{eq:xtol}.}

Figure~\ref{fig:xmatrix} shows the $\mathbf{X}-\mathbf{I}_4$ matrix elements for the error introduced by the change in switching angle for a $0.5~\mathrm{K}$ change in temperature.
The $Q$-to-$U$ term is the worst offender and is just below the limit at $500~\mathrm{nm}$.

There are other contributors to $\mathbf{X}$ than changes in switching angle with temperature.
E.g., the retardances of the \edit1{FLCs} and the retarder also have a small temperature dependence.
The polarimetric calibration procedure also has a finite accuracy \citep{2008A&A...489..429V}.
We do not explicitly model these effects here, since the switching angle is expected to be the dominant source of error.
\edit1{For example, polycarbonate retarders have typical temperature coefficients around $0.4~\mathrm{pm}/\mathrm{nm}/\mathrm{K}$, and a $5~\mathrm{K}$ change in temperature is required to exceed the permissible error.}
Instead we assign a fraction of the permissible error to changes in switching angle and set the requirement for thermal stability to $\pm0.2~\mathrm{K}$.

\section{Opto-Mechanical Design}

Figure~\ref{fig:crosssection} shows a cross-section of the modulator.
The mechanical design borrows heavily from the HAO Lyot filter designs used in the CoMP, CoMP-S, SCD, and ChroMag instruments.
The modulator optics are glued into mounts that allow the optic to be oriented to any angle using a RTV silicone.
The mounts consist of two parts.
The inner part is round and holds the optic.
It can be oriented to the desired angle and glued to the hexagonal outer part that is indexed to the inner mount assembly.
The optics stack is assembled between parallel windows using index-matching gel.
The windows rest on O-rings in their mounts.
The entrance window mount is spring-loaded against the inner mount assembly with 4.4~N.

\narrowfig{"f4"}{fig:crosssection}{A cross-section view of the modulator with major components labeled.
\edit1{Optical components are shown in dark blue:}
1.~entrance window;
2.~\edit1{FLC} 1;
3.~\edit1{FLC} 2;
4.~fixed retarder; and
5.~exit window.
\edit1{The mechanical assembly is color-coded by its major components:}
6.~pressure plate \edit1{(dark green)};
7.~oven \edit1{(gray)};
8.~optic holders \edit1{(yellow)};
9.~inner mount assembly \edit1{(light green)}; and
10.~Delrin shell \edit1{(light blue)}.
See the text for details on the mechanical design.}

The inner mount assembly is inserted in an oven consisting of an aluminum tube with a silicone rubber heater element and aerogel insulation wrapped around it.
An off-the-shelf precision temperature controller is used to stabilize the oven to $35.0\degr\textrm{C}$ to better than $0.1\degr\textrm{C}$.
The modulator is encased in a Delrin housing.
Electrical connections for the \edit1{FLCs} and the heater system are routed to two D-subminiature connectors on the housing.

A custom controller based on an Arduino Uno microcontroller board was built to drive the \edit1{FLCs}.
The camera software sends a voltage sequence to the controller via a serial interface, which is preloaded into two Burr-Brown DAC714 digital-to-analog converters (DACs).
A synchronization pulse is then used to update the voltages when the chopper that controls the exposure of the cameras is closed.
The \edit1{FLCs} are primarily capacitive loads, with capacitance of about $80~\mathrm{nF}$.
The DACs are capable of driving $5~\mathrm{mA}$, which is more than adequate to drive the \edit1{FLCs} between states in under a millisecond.
The controller also resets the voltage to zero after a few seconds of inactivity as a safety feature because the \edit1{FLCs} may be damaged if driven by a constant voltage for a prolonged period of time.

\section{Performance}\label{sec:performance}

The components of the modulator must be accurately aligned to ensure proper functioning of the assembled device.
HAO has a facility Lab Spectropolarimeter (LSPM) test setup for polarimetric characterization of optics that was used to test components of the CRISP modulator after they were mounted.

\narrowfig{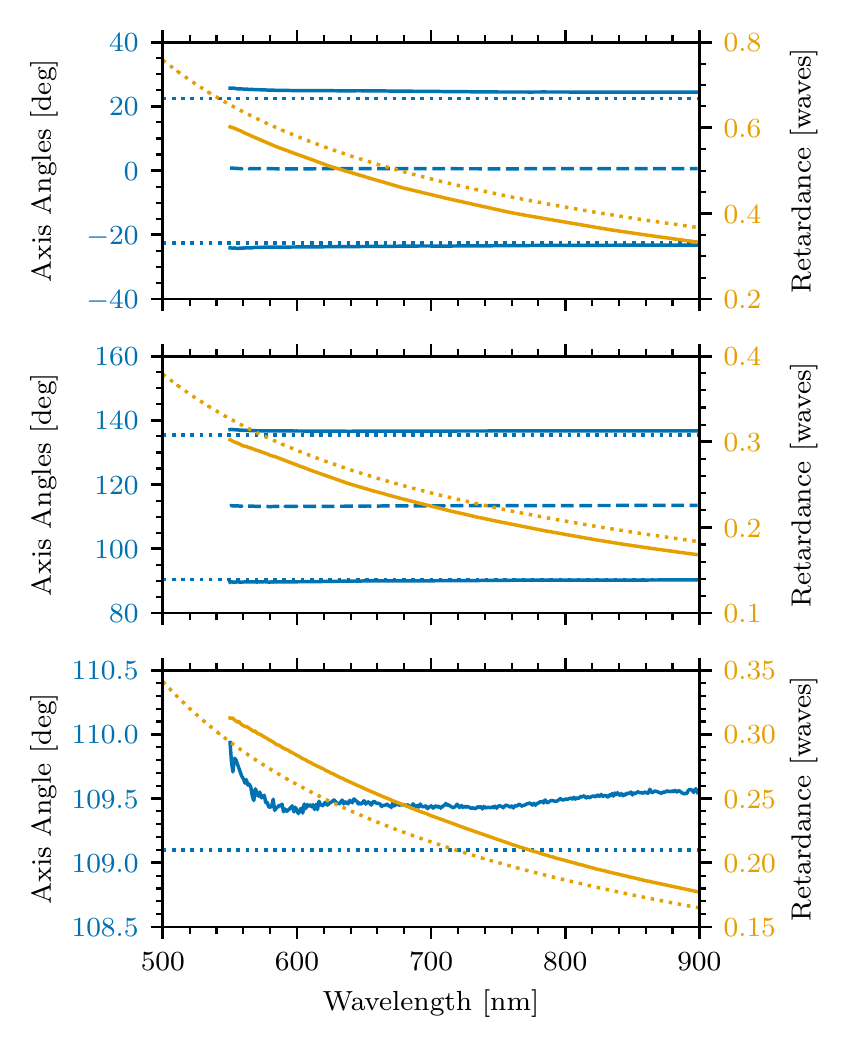}{fig:LSPM_tests}{Solid lines: retardances (orange) and fast axis positions (blue) of the $\frac12$-wave \edit1{FLC} (top panel), the $\frac14$-wave \edit1{FLC} (middle panel), and the retarder (bottom panel), as determined by a fit to the Mueller matrix inferred from LSPM measurements at room temperature.
	The panels for the \edit1{FLCs} show two fast axis positions for $\pm8~\mathrm{V}$ drive voltages.
	Dashed lines: fast axis bisector.
	Dotted lines: design retardances and fast axis positions.
In these measurements the signal level drops below acceptable levels around $550~\mathrm{nm}$.}

The LSPM consists of relay optics that feed light from a halogen bulb through, in order, a calibration package that consists of a polarizer and a retarder in individual rotation stages, the sample under test, and a polychromatic polarization modulator and analyzer, into an Ocean Optics USB4000 fiber-fed spectrograph.
This setup allows for characterization of the full Mueller matrix of the sample as a function of wavelength.
The spectrograph covers the wavelength range from about $450~\textrm{nm}$ to about $1100~\textrm{nm}$, though signal levels are low under $550~\textrm{nm}$.

We solve for retardance and fast axis position of a linear retarder that matches the Mueller matrices derived from LSPM measurements as a function of wavelength.
Figure~\ref{fig:LSPM_tests} shows the results for the three CRISP modulator components.
The figure also shows the design retardances and fast axis positions.
The $\lambda/2$ and $\lambda/4$ \edit1{FLCs} have measured retardances of $0.47\lambda$ and $0.24\lambda$ at $665~\mathrm{nm}$, \edit1{and mean switching angles of $48.2\degr$ and $46.7\degr$.}
The fixed retarder is measured at $0.251\lambda$.
\edit1{All three components show a curious increase in the fast axis position at the shortest wavelengths.
However, the signal level is low, and it may be that the effect is the result of systematic errors in the measurement.}

As discussed in Sect.~\ref{sec:design}, we can re-optimize the design with these values.
However, we made our measurements at room temperature.
The measurement should have been performed at the operating temperature of $35\degr\mathrm{C}$ because component retardance has some temperature dependence.
Using the retardance values at room temperature we find fast axis angles for the 2nd \edit1{FLC} and the fixed retarder are $113.5$ and $108.5~\mathrm{degrees}$.
However, if we assume the retarder will have its design retardance at $35\degr\mathrm{C}$, the fast axis positions revert to the nominal design.
\edit1{The FLC switching angles are larger than the nominal $45\degr$ design, but also expected to decrease at a higher operating temperature, possibly to angles below $45\degr$ \citep{2003SPIE.4843...45G,Gisler2005}.}
We choose not to change the modulator design because of the unknown effect of temperature on the component retardance \edit1{and FLC switching angle,} and because only marginal improvement of performance is expected.

The modulator was first assembled with air gaps, and once the proper relative alignment of the components was confirmed using the LSPM, the modulator was assembled in its housing using Nye OCF-452 optical coupling fluid on the glass interfaces.
The purpose of the coupling fluid is to reduce internal Fresnel reflections between the surfaces of the optics.
Nye OCF-452 was used because it has a refractive index that is well-matched to the Corning XG glass of the \edit1{FLCs} and the BK7 glass of the retarder and windows.
Internal Fresnel reflections at the optical interfaces of the components are limited to below the $2\times10^{-5}$ level.

\narrowfig{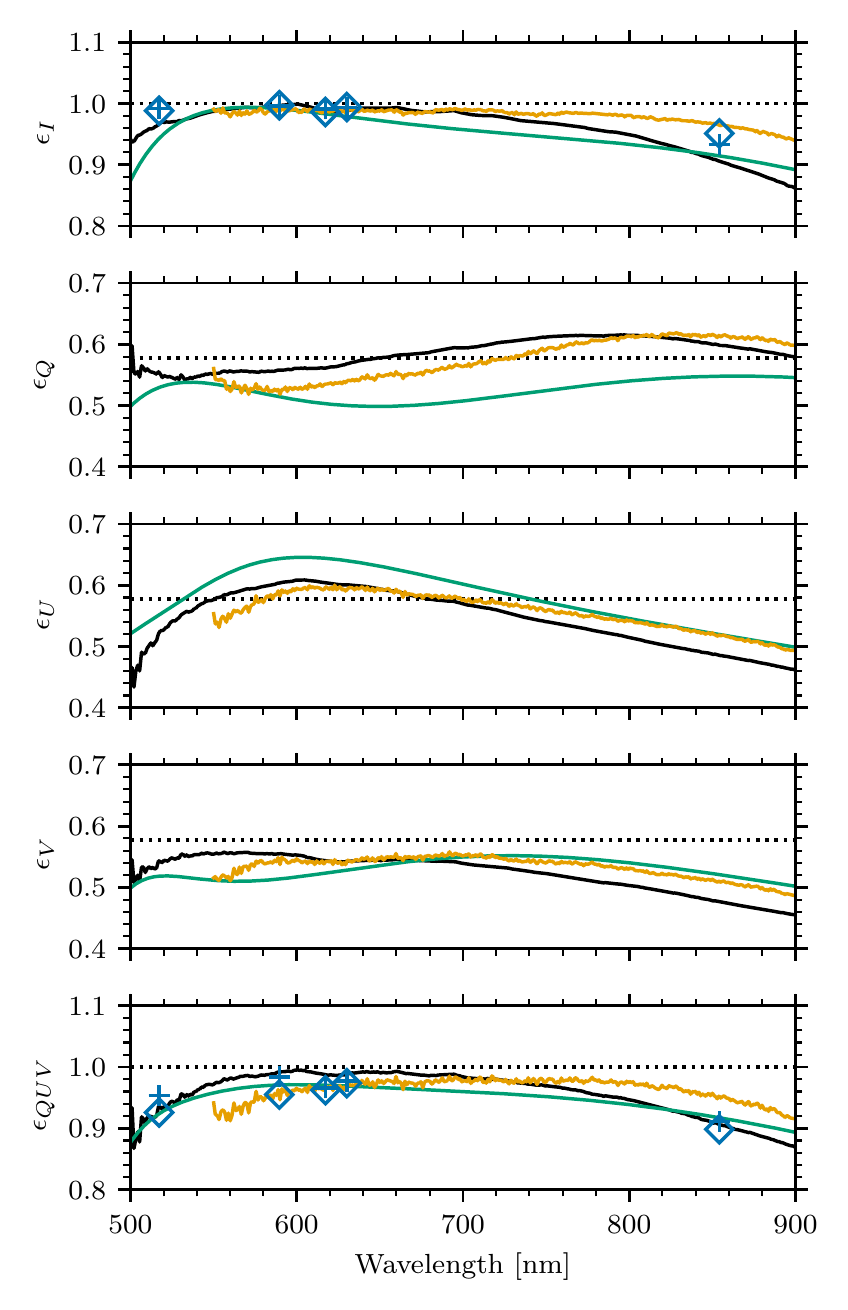}{fig:eff_wet}{\edit1{Modulation efficiencies of the modulator.
The top four panels shows the efficiency in Stokes $I$, $Q$, $U$, and $V$.
The bottom panel shows the RSS of the efficiencies in Stokes $Q$, $U$, and $V$.}
Solid black lines: modulation efficiencies of the assembled modulator measured with the LSPM.
\edit1{Solid green lines: design efficiencies.}
Solid orange lines: modulation efficiencies expected from the measurements in Fig.~\ref{fig:LSPM_tests}.
Blue crosses and diamonds: modulation efficiencies measured at the telescope in the transmitted and reflected beams, resp.
The horizontal dotted line in each panel shows the theoretical maximum efficiency for a balanced modulation scheme.}

The fully assembled modulator was brought to operating temperature and tested again on the LSPM.
The LSPM produces measurements of the Mueller matrix of the modulator in each of its 4 states.
We simulate a perfect analyzer in $Q$ to calculate modulation efficiencies, shown as a function of wavelength in Fig.~\ref{fig:eff_wet}.

The measured efficiencies largely show the expected behavior when compared to the design \edit1{shown in solid green curves in Fig.~\ref{fig:eff_wet}}.
However, differences in the model and measured efficiencies cannot be fully attributed to as-built retardances and component alignment.
The differences are likely due to several factors that were not included in the tolerance analysis.
The model assumes that the components are perfect retarders with a known dispersion of birefringence, and that the \edit1{FLCs} have an exact $45\degr$ switching angle.
In reality, the components have imperfections such as chromatic variation of the fast axis, and the actual dispersion of birefringence is different from the model.
This can be seen in Fig.~\ref{fig:LSPM_tests}.
The solid blue lines are not horizontal, and the solid and dotted orange lines are not parallel.
The \edit1{FLC} switching angle is also not exactly $45\degr$.
Lastly, bulk rotational alignment of the modulator to the analyzer was not included in the analysis.
\edit1{The efficiencies of some} modulator designs, in particular the traditional rotating retarder, are invariant under rotation of the modulator with respect to the analyzer.
This design is not invariant, and rotation of the modulator results in depressed efficiencies.

Figure~\ref{fig:eff_wet} also shows the modulation efficiencies computed from the Mueller matrices of the measured components \edit1{in orange}.
They show good agreement with the measured efficiencies of the assembled modulator.
We attribute the differences mostly to the components not being measured at operating temperature.
There are also likely small differences in the relative orientation of the components in the assembled modulator compared to the individual measurements.

The CRISP instrument is intended for high-resolution imaging.
The modulator, therefore, must have low transmitted wavefront distortion (TWD).
Because the internal optics are coupled using an index-matching gel, the TWD is dominated by the by the entrance and exit windows.
Fortunately, excellent quality windows are inexpensive and commonly available.
The TWD of the assembled modulator was measured using a Zygo interferometer.
It was found to be $0.20~\mathrm{waves}$ at $632~\mathrm{nm}$ RMS over the clear aperture after removal of the tilt component, but including $0.14~\mathrm{waves}$ of power that introduces primarily a shift in focus position.

\begin{table}[tbp]
\caption{\label{tab:efficiencies}Modulation efficiencies measured at the telescope averaged over the field of view.}
\centering
\begin{tabular}{l*{5}{>{$}c<{$}}}
\hline
Wavelength & \epsilon_I & \epsilon_Q & \epsilon_U & \epsilon_V & \epsilon_{QUV}\\ \hline\hline
\multicolumn{6}{c}{Transmitted} \\\hline
$517.3~\mathrm{nm}$ & 0.992&0.500&0.588&0.560&0.954\\
$589.6~\mathrm{nm}$ & 0.996&0.573&0.591&0.539&0.984\\
$617.3~\mathrm{nm}$ & 0.991&0.506&0.608&0.554&0.966\\
$630.2~\mathrm{nm}$ & 0.993&0.576&0.579&0.536&0.977\\
$854.2~\mathrm{nm}$ & 0.933&0.570&0.504&0.501&0.911\\\hline\hline
\multicolumn{6}{c}{Reflected} \\\hline
$517.3~\mathrm{nm}$ & 0.988&0.485&0.572&0.543&0.926\\
$589.6~\mathrm{nm}$ & 0.997&0.557&0.572&0.524&0.955\\
$617.3~\mathrm{nm}$ & 0.986&0.506&0.607&0.549&0.962\\
$630.2~\mathrm{nm}$ & 0.994&0.576&0.579&0.530&0.974\\
$854.2~\mathrm{nm}$ & 0.951&0.563&0.497&0.493&0.898\\
\end{tabular}
\end{table}

The modulator was installed at the SST in October 2014.
\edit1{CRISP uses a polarizing beamsplitter to analyze the polarization signal in two orthogonal directions simultaneously \citep{2015A&A...573A..40D}.
This kind of setup is known as a dual-beam polarimeter and allows for the removal of crosstalk resulting from atmospheric seeing from Stokes $I$ to Stokes $Q$, $U$, and $V$ \citep{2012ApJ...757...45C}.}
\edit1{Measured modulation efficiency averaged over the field of view for the transmitted and reflected beams at five wavelengths commonly used for solar polarimetry are given in Table~\ref{tab:efficiencies} and also shown in Fig.~\ref{fig:eff_wet} as blue crosses and diamonds.}

\edit1{It is not possible to directly compare the efficiencies in the individual Stokes parameters.}
The telescope measurements include all the optics on the tables, which include a number of mirrors and lenses, a dichroic beamsplitter, the CRISP prefilter, a gray beamsplitter, and the CRISP etalons.
These elements cannot be separated from the modulator.
The calibration procedure fits all optics on the table between the calibration optics and the polarization analyzer as one modulation matrix \citep{2008A&A...489..429V}.
In effect, all optical elements between the calibration optics and the polarimetric analyzer together act as the modulator.
\edit1{There are oblique reflections that may cause some mixing of all Stokes parameters due to retardance of the mirror coatings, and the reference frame of polarization of these measurements is not normal to the optical table.
It is still valuable to compare the performance in the telescope to the design performance and lab measurements, but this can only be done in an aggregate way, i.e., by comparing the RSS of the efficiencies in Stokes $Q$, $U$, and $V$,
\begin{equation}
	\epsilon_{QUV} = \sqrt{\epsilon_Q^2+\epsilon_U^2+\epsilon_V^2}.
\end{equation}
The overall agreement of the performance in the telescope setup and the assembled modulator is very good.
The largest difference in the RSS of the efficiencies in $Q$, $U$, and $V$ is $3.7\%$ in the reflected beam at $589.6~\mathrm{nm}$.}

The transmitted and reflected beams show very similar behavior.
\edit1{The differences in efficiencies between the beams are on the order of a few percent, and may result from differences in the contrast of the  polarizing beamsplitter in the transmitted and reflected beams, or from polarizing components in the telescope such as the wide-band beamsplitter that has a highly uneven ratio of the transmitted and reflected light.}

The overall performance is excellent with the lowest efficiencies only slightly below $50\%$ (cf.~the optimum and balanced efficiency of $57.7\%$).
\edit1{The RSS of the efficiencies in Stokes $Q$, $U$, and $V$ are $94\%$, $97\%$, $96\%$, $98\%$, and $90\%$ for these 5 wavelengths.}

\section{Conclusion}\label{sec:conclusion}

The trade-offs and procedures described in this paper were employed to design the polarimetric modulator for the CRISP instrument, but can be applied to the design of modulators for other instruments.
We chose to omit some steps that could result in somewhat improved performance of the modulator.
If schedule permits, it is possible to incrementally optimize the design with measured optic properties.
We could have delayed the purchase of the retarder until the \edit1{FLCs}, which have the largest errors, had been characterized at operating temperature, so that the value of the retarder could have been optimized for the as-built \edit1{FLCs}.
While this design with only three components is robust, such incremental re-optimization may be necessary to guarantee acceptable efficiencies for modulator designs with more optical elements that cover larger wavelength ranges \citep{2012SPIE.8446E..25S}.

We did not specifically consider polarized spectral ``fringes'' in our design process.
A description of polarized spectral fringes can be found in reviews by \cite{1991sopo.work..166L}, \cite{2003A&A...401....1S}, and \cite{2004JOptA...6.1036C}.
They are interference patterns that are produced by reflections between parallel surfaces in a system with polarization optics, such as the components of the modulator, that are difficult to characterize \citep{2017JATIS...3d8001H} and remove \citep{2006ApJ...649..553R,2012ApJ...756..194C,2019ApJ...872..173C}.
\cite{2015SPIE.9613E..0GS} optimized components to suppress polarized fringes for their application by ensuring that the periods of the fringes are much smaller than the spectral resolution of their instrument.
\edit1{This approach cannot be applied in modulators using FLCs (or LCVRs) because the FLC layer thickness that determines the period of the fringes is set by the required retardance.
Because the FLC layer is very thin, fringes caused by reflections at the FLC-glass interfaces have periods of ten or more of nanometers \citep{2003SPIE.4843...45G}, i.e., much larger than the CRISP bandpass of less than ten picometers \citep{2008ApJ...689L..69S}.
These fringes will consequently be relatively stable over the CRISP bandpass and therefore will be nearly completely removed in the polarimetric calibration.
Fringes caused by reflections between surfaces at larger optical distance have smaller periods and can be a problem.
For those,} the only available option \edit1{is to reduce the amplitude of fringes by reducing the amplitude of} Fresnel reflections from the interfaces of the optical elements.
The use of optical coupling fluid is therefore not only required to address etaloning, but also to suppress these fringes.

\widefig{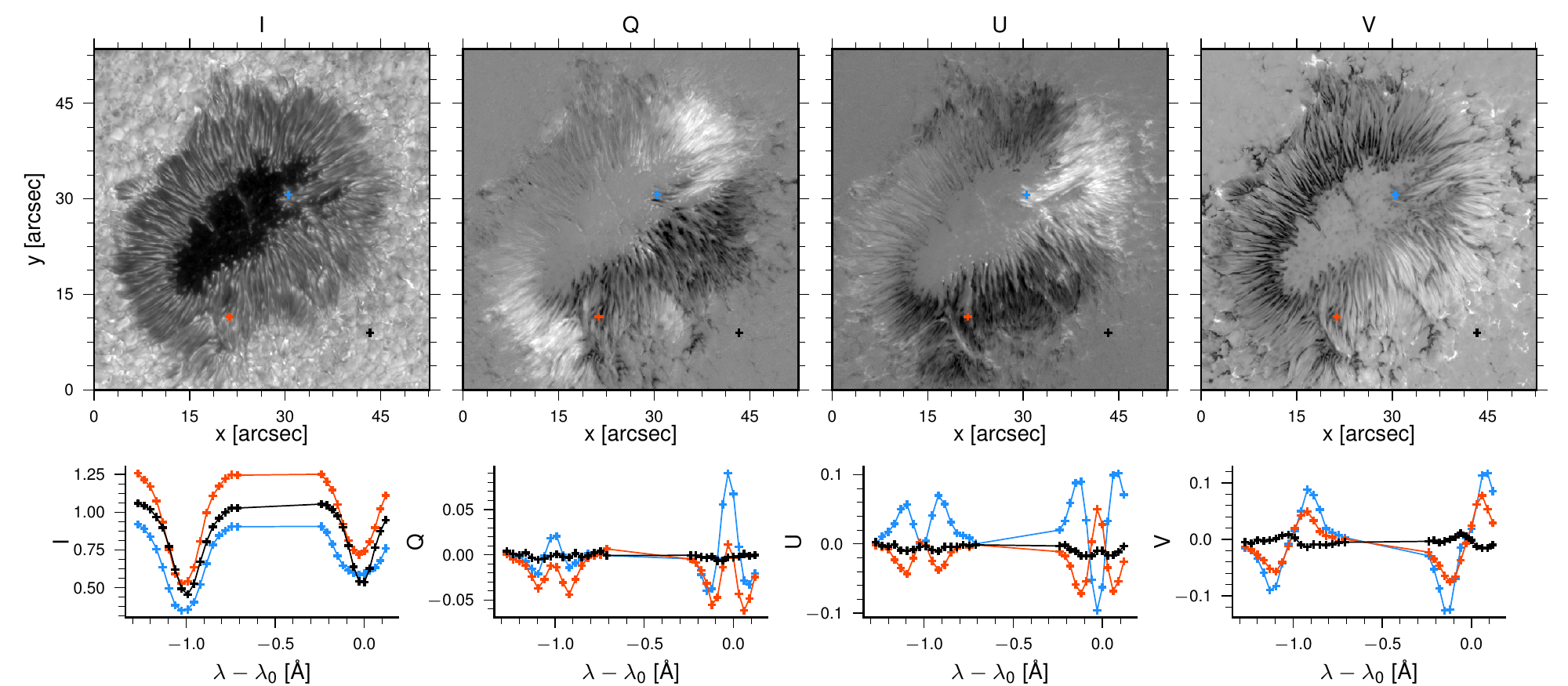}{fig:exdata}{Example spectro-polarimetric data of AR12471 in the \ion{Fe}{1} lines at $630.2~\mathrm{nm}$.
The observation was recorded on 2019-05-10 around 09:10~UT.
	Top row: images of the Stokes parameter in the blue wing of the $630.25~\mathrm{nm}$ line at $-90~\mathrm{mA}$ from line center.
	Spectra at the black, red, and blue crosses are shown in the bottom row.
The locations of the spectral sampling are indicated by crosses.}

The polarimetric modulator described here has been in use for science observations at the SST starting with the 2015 observing season.
Example data of a sunspot are shown in Fig.~\ref{fig:exdata}.
The data reduction procedures are described in detail by \cite{2015A&A...573A..40D} and \cite{2018arXiv180403030L}.
These data can be fit using forward-modeling procedures to derive quantitative measures of atmospheric parameters, most notably the strength and direction of magnetic field.
For example, \cite{2018ApJ...860...10K} studied the structure and evolution of temperature and magnetic field in a flaring active region using full-Stokes CRISP observations in the \ion{Ca}{2} line at $854.2~\mathrm{nm}$, \cite{2019A&A...627A.101V} used similar data in combination with data from the IRIS mission \citep{2014SoPh..289.2733D} to study Ellerman bombs and UV bursts, \cite{2020arXiv200901537V} inferred the photopheric and chromospheric magnetic field vector in a flare target and studied their differences, \cite{2019A&A...621A..35L} used CRISP observations in the \ion{He}{1} $\mathrm{D}_3$ line in a study of a flare, \cite{2020arXiv200614487M} and \cite{2020arXiv200614486P} studied chromospheric magnetic fields in plage targets and estimated a canopy mean field strength of $400~\mathrm{G}$ in the chromosphere, and \cite{2020A&A...641L...5J} studied very small-scale reconnection in the solar photosphere using CRISP polarimetry and CHROMIS \citep{2019A&A...626A..55S} observations in H$\beta$.

Figure~\ref{fig:qs} shows a region of quiet sun with the line-of-sight component of the magnetic field inferred from full-Stokes observations of the \ion{Fe}{1} $617.3~\mathrm{nm}$ line profile using a spatially-regularized Milne-Eddington inversion method \citep{2019A&A...631A.153D}.
This example highlights the power of CRISP combined with this modulator.
Quiet-sun magnetic fields are weak and difficult to detect.
Telescopes and instruments that achieve high spatial resolution, have adequate spectral resolving power, and have high system efficiency are required to study them.
We refer the interested reader to \cite{2019LRSP...16....1B} for a comprehensive review of observations of quiet-sun magnetic field.

\narrowfig{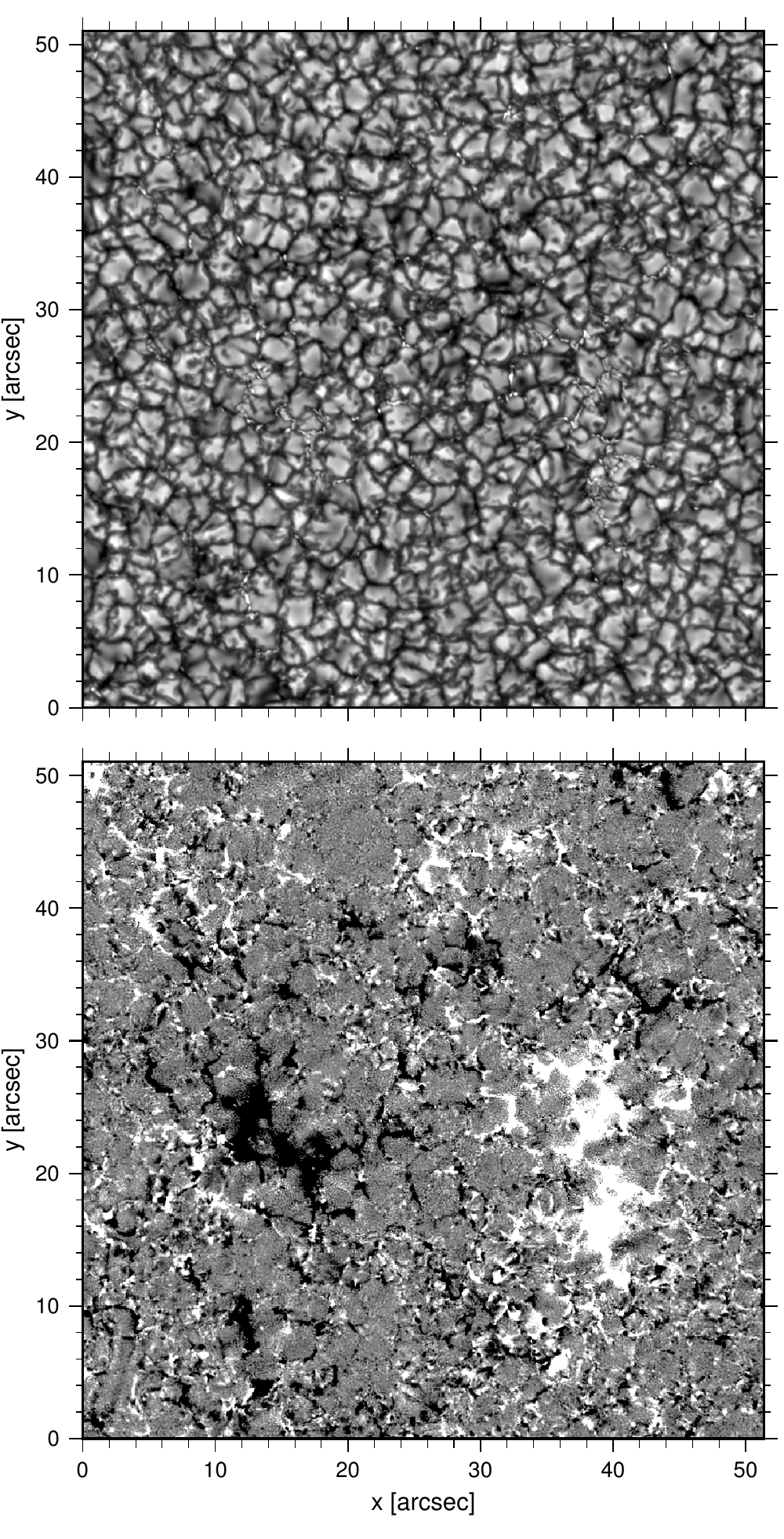}{fig:qs}{Intensity and line-of-sight magnetic field strength for a quiet sun region observed on 2020-07-14 around 08:41~UT during a period of very good seeing conditions.
The magnetic field strength was inferred from observations of the \ion{Fe}{1} $617.3~\mathrm{nm}$ line and is scaled between $-25$ and $25~\mathrm{G}$.}

The high throughput and efficiency of CRISP with this modulator also enables observations in many lines with polarimetry while maintaining sufficient cadence for studies of dynamic events.
Such multi-line observations were used by \cite{2018A&A...612A..28L} in a study of chromospheric heating in an emerging flux region.
They used the STiC code \citep{2019A&A...623A..74D} to simultaneously interpret the signals from several lines.
\cite{2019ApJ...870...88E} used the same code in a similar way to study penumbral microjets.

\acknowledgments{We acknowledge R.~Casini for the development of the codes used for optimization and tolerancing of the modulator designs.
This material is based upon work supported by the National Center for Atmospheric Research, which is a major facility sponsored by the National Science Foundation under Cooperative Agreement No. 1852977.
CRISP and the modulator were funded by the Marianne and Marcus Wallenberg Foundation.
This research has made use of NASA's Astrophysics Data System,
NumPy \citep{2011CSE....13b..22V},
matplotlib, a Python library for publication quality graphics \citep{2007CSE.....9...90H},
Astropy, a community-developed core Python package for Astronomy \citep{2018AJ....156..123A, 2013A&A...558A..33A},
and the IPython package \citep{2007CSE.....9c..21P}.
The acknowledgements were compiled using the Astronomy Acknowledgement Generator.}


\begin{thebibliography}{59}
\expandafter\ifx\csname natexlab\endcsname\relax\def\natexlab#1{#1}\fi

\bibitem[{{Astropy Collaboration} {et~al.}(2018){Astropy Collaboration},
  {Price-Whelan}, {Sip{\H{o}}cz}, {G{\"u}nther}, {Lim}, {Crawford}, {Conseil},
  {Shupe}, {Craig}, {Dencheva}, {Ginsburg}, {Vand erPlas}, {Bradley},
  {P{\'e}rez-Su{\'a}rez}, {de Val-Borro}, {Aldcroft}, {Cruz}, {Robitaille},
  {Tollerud}, {Ardelean}, {Babej}, {Bach}, {Bachetti}, {Bakanov}, {Bamford},
  {Barentsen}, {Barmby}, {Baumbach}, {Berry}, {Biscani}, {Boquien}, {Bostroem},
  {Bouma}, {Brammer}, {Bray}, {Breytenbach}, {Buddelmeijer}, {Burke},
  {Calderone}, {Cano Rodr{\'\i}guez}, {Cara}, {Cardoso}, {Cheedella}, {Copin},
  {Corrales}, {Crichton}, {D'Avella}, {Deil}, {Depagne}, {Dietrich}, {Donath},
  {Droettboom}, {Earl}, {Erben}, {Fabbro}, {Ferreira}, {Finethy}, {Fox},
  {Garrison}, {Gibbons}, {Goldstein}, {Gommers}, {Greco}, {Greenfield},
  {Groener}, {Grollier}, {Hagen}, {Hirst}, {Homeier}, {Horton}, {Hosseinzadeh},
  {Hu}, {Hunkeler}, {Ivezi{\'c}}, {Jain}, {Jenness}, {Kanarek}, {Kendrew},
  {Kern}, {Kerzendorf}, {Khvalko}, {King}, {Kirkby}, {Kulkarni}, {Kumar},
  {Lee}, {Lenz}, {Littlefair}, {Ma}, {Macleod}, {Mastropietro}, {McCully},
  {Montagnac}, {Morris}, {Mueller}, {Mumford}, {Muna}, {Murphy}, {Nelson},
  {Nguyen}, {Ninan}, {N{\"o}the}, {Ogaz}, {Oh}, {Parejko}, {Parley}, {Pascual},
  {Patil}, {Patil}, {Plunkett}, {Prochaska}, {Rastogi}, {Reddy Janga},
  {Sabater}, {Sakurikar}, {Seifert}, {Sherbert}, {Sherwood-Taylor}, {Shih},
  {Sick}, {Silbiger}, {Singanamalla}, {Singer}, {Sladen}, {Sooley},
  {Sornarajah}, {Streicher}, {Teuben}, {Thomas}, {Tremblay}, {Turner},
  {Terr{\'o}n}, {van Kerkwijk}, {de la Vega}, {Watkins}, {Weaver}, {Whitmore},
  {Woillez}, {Zabalza}, \& {Astropy Contributors}}]{2018AJ....156..123A}
{Astropy Collaboration}, {Price-Whelan}, A.~M., {Sip{\H{o}}cz}, B.~M., {et~al.}
  2018, \aj, 156, 123

\bibitem[{{Astropy Collaboration} {et~al.}(2013){Astropy Collaboration},
  {Robitaille}, {Tollerud}, {Greenfield}, {Droettboom}, {Bray}, {Aldcroft},
  {Davis}, {Ginsburg}, {Price-Whelan}, {Kerzendorf}, {Conley}, {Crighton},
  {Barbary}, {Muna}, {Ferguson}, {Grollier}, {Parikh}, {Nair}, {Unther},
  {Deil}, {Woillez}, {Conseil}, {Kramer}, {Turner}, {Singer}, {Fox}, {Weaver},
  {Zabalza}, {Edwards}, {Azalee Bostroem}, {Burke}, {Casey}, {Crawford},
  {Dencheva}, {Ely}, {Jenness}, {Labrie}, {Lim}, {Pierfederici}, {Pontzen},
  {Ptak}, {Refsdal}, {Servillat}, \& {Streicher}}]{2013A&A...558A..33A}
{Astropy Collaboration}, {Robitaille}, T.~P., {Tollerud}, E.~J., {et~al.} 2013,
  \aap, 558, A33

\bibitem[{{Bellot Rubio} \& {Orozco Su{\'a}rez}(2019)}]{2019LRSP...16....1B}
{Bellot Rubio}, L. \& {Orozco Su{\'a}rez}, D. 2019, Living Reviews in Solar
  Physics, 16, 1

\bibitem[{{Casini} {et~al.}(2012{\natexlab{a}}){Casini}, {de Wijn}, \&
  {Judge}}]{2012ApJ...757...45C}
{Casini}, R., {de Wijn}, A.~G., \& {Judge}, P.~G. 2012{\natexlab{a}}, \apj,
  757, 45

\bibitem[{{Casini} {et~al.}(2012{\natexlab{b}}){Casini}, {Judge}, \&
  {Schad}}]{2012ApJ...756..194C}
{Casini}, R., {Judge}, P.~G., \& {Schad}, T.~A. 2012{\natexlab{b}}, \apj, 756,
  194

\bibitem[{{Casini} \& {Li}(2019)}]{2019ApJ...872..173C}
{Casini}, R. \& {Li}, W. 2019, \apj, 872, 173

\bibitem[{{Clarke}(2004)}]{2004JOptA...6.1036C}
{Clarke}, D. 2004, Journal of Optics A: Pure and Applied Optics, 6, 1036

\bibitem[{{de la Cruz Rodr{\'\i}guez}(2019)}]{2019A&A...631A.153D}
{de la Cruz Rodr{\'\i}guez}, J. 2019, \aap, 631, A153

\bibitem[{{de la Cruz Rodr{\'\i}guez} {et~al.}(2019){de la Cruz
  Rodr{\'\i}guez}, {Leenaarts}, {Danilovic}, \&
  {Uitenbroek}}]{2019A&A...623A..74D}
{de la Cruz Rodr{\'\i}guez}, J., {Leenaarts}, J., {Danilovic}, S., \&
  {Uitenbroek}, H. 2019, \aap, 623, A74

\bibitem[{{de la Cruz Rodr{\'\i}guez} {et~al.}(2015){de la Cruz
  Rodr{\'\i}guez}, {L{\"o}fdahl}, {S{\"u}tterlin}, {Hillberg}, \& {Rouppe van
  der Voort}}]{2015A&A...573A..40D}
{de la Cruz Rodr{\'\i}guez}, J., {L{\"o}fdahl}, M.~G., {S{\"u}tterlin}, P.,
  {Hillberg}, T., \& {Rouppe van der Voort}, L. 2015, \aap, 573, A40

\bibitem[{{De Pontieu} {et~al.}(2014){De Pontieu}, {Title}, {Lemen}, {Kushner},
  {Akin}, {Allard}, {Berger}, {Boerner}, {Cheung}, {Chou}, {Drake}, {Duncan},
  {Freeland}, {Heyman}, {Hoffman}, {Hurlburt}, {Lindgren}, {Mathur}, {Rehse},
  {Sabolish}, {Seguin}, {Schrijver}, {Tarbell}, {W{\"u}lser}, {Wolfson},
  {Yanari}, {Mudge}, {Nguyen-Phuc}, {Timmons}, {van Bezooijen}, {Weingrod},
  {Brookner}, {Butcher}, {Dougherty}, {Eder}, {Knagenhjelm}, {Larsen},
  {Mansir}, {Phan}, {Boyle}, {Cheimets}, {DeLuca}, {Golub}, {Gates}, {Hertz},
  {McKillop}, {Park}, {Perry}, {Podgorski}, {Reeves}, {Saar}, {Testa}, {Tian},
  {Weber}, {Dunn}, {Eccles}, {Jaeggli}, {Kankelborg}, {Mashburn}, {Pust},
  {Springer}, {Carvalho}, {Kleint}, {Marmie}, {Mazmanian}, {Pereira}, {Sawyer},
  {Strong}, {Worden}, {Carlsson}, {Hansteen}, {Leenaarts}, {Wiesmann},
  {Aloise}, {Chu}, {Bush}, {Scherrer}, {Brekke}, {Martinez-Sykora}, {Lites},
  {McIntosh}, {Uitenbroek}, {Okamoto}, {Gummin}, {Auker}, {Jerram}, {Pool}, \&
  {Waltham}}]{2014SoPh..289.2733D}
{De Pontieu}, B., {Title}, A.~M., {Lemen}, J.~R., {et~al.} 2014, \solphys, 289,
  2733

\bibitem[{{de Wijn} {et~al.}(2012){de Wijn}, {Bethge}, {Tomczyk}, \&
  {McIntosh}}]{2012SPIE.8446E..78D}
{de Wijn}, A.~G., {Bethge}, C., {Tomczyk}, S., \& {McIntosh}, S. 2012, in
  Society of Photo-Optical Instrumentation Engineers (SPIE) Conference Series,
  Vol. 8446, Ground-based and Airborne Instrumentation for Astronomy IV, 844678

\bibitem[{{del Toro Iniesta} \& {Collados}(2000)}]{2000ApOpt..39.1637D}
{del Toro Iniesta}, J.~C. \& {Collados}, M. 2000, \ao, 39, 1637

\bibitem[{{Elmore} {et~al.}(2008){Elmore}, {Casini}, {Card}, {Davis},
  {Lecinski}, {Lull}, {Nelson}, \& {Tomczyk}}]{2008SPIE.7014E..16E}
{Elmore}, D.~F., {Casini}, R., {Card}, G.~L., {et~al.} 2008, in Society of
  Photo-Optical Instrumentation Engineers (SPIE) Conference Series, Vol. 7014,
  Ground-based and Airborne Instrumentation for Astronomy II, 701416

\bibitem[{{Esteban Pozuelo} {et~al.}(2019){Esteban Pozuelo}, {de la Cruz
  Rodr{\'\i}guez}, {Drews}, {Rouppe van der Voort}, {Scharmer}, \&
  {Carlsson}}]{2019ApJ...870...88E}
{Esteban Pozuelo}, S., {de la Cruz Rodr{\'\i}guez}, J., {Drews}, A., {et~al.}
  2019, \apj, 870, 88

\bibitem[{{Gandorfer}(1999)}]{1999OptEn..38.1402G}
{Gandorfer}, A.~M. 1999, Optical Engineering, 38, 1402

\bibitem[{{Gisler}(2005)}]{Gisler2005}
{Gisler}, D. 2005, PhD thesis, ETH Zurich

\bibitem[{{Gisler} {et~al.}(2003){Gisler}, {Feller}, \&
  {Gandorfer}}]{2003SPIE.4843...45G}
{Gisler}, D., {Feller}, A., \& {Gandorfer}, A.~M. 2003, in Society of
  Photo-Optical Instrumentation Engineers (SPIE) Conference Series, Vol. 4843,
  \procspie, ed. S.~{Fineschi}, 45--54

\bibitem[{{Harrington} {et~al.}(2017){Harrington}, {Snik}, {Keller}, {Sueoka},
  \& {van Harten}}]{2017JATIS...3d8001H}
{Harrington}, D.~M., {Snik}, F., {Keller}, C.~U., {Sueoka}, S.~R., \& {van
  Harten}, G. 2017, Journal of Astronomical Telescopes, Instruments, and
  Systems, 3, 048001

\bibitem[{{Harrington} \& {Sueoka}(2018)}]{2018JATIS...4d4006H}
{Harrington}, D.~M. \& {Sueoka}, S.~R. 2018, Journal of Astronomical
  Telescopes, Instruments, and Systems, 4, 044006

\bibitem[{{Hunter}(2007)}]{2007CSE.....9...90H}
{Hunter}, J.~D. 2007, Computing in Science and Engineering, 9, 90

\bibitem[{{Ichimoto} {et~al.}(2008){Ichimoto}, {Lites}, {Elmore}, {Suematsu},
  {Tsuneta}, {Katsukawa}, {Shimizu}, {Shine}, {Tarbell}, {Title}, {Kiyohara},
  {Shinoda}, {Card}, {Lecinski}, {Streander}, {Nakagiri}, {Miyashita},
  {Noguchi}, {Hoffmann}, \& {Cruz}}]{2008SoPh..249..233I}
{Ichimoto}, K., {Lites}, B., {Elmore}, D., {et~al.} 2008, \solphys, 249, 233

\bibitem[{{Iglesias} \& {Feller}(2019)}]{2019OptEn..58h2417I}
{Iglesias}, F.~A. \& {Feller}, A. 2019, Optical Engineering, 58, 082417

\bibitem[{{Iglesias} {et~al.}(2016){Iglesias}, {Feller}, {Nagaraju}, \&
  {Solanki}}]{2016A&A...590A..89I}
{Iglesias}, F.~A., {Feller}, A., {Nagaraju}, K., \& {Solanki}, S.~K. 2016,
  \aap, 590, A89

\bibitem[{{Joshi} {et~al.}(2020){Joshi}, {Rouppe van der Voort}, \& {de la Cruz
  Rodr{\'\i}guez}}]{2020A&A...641L...5J}
{Joshi}, J., {Rouppe van der Voort}, L. H.~M., \& {de la Cruz Rodr{\'\i}guez},
  J. 2020, \aap, 641, L5

\bibitem[{{Judge} {et~al.}(2004){Judge}, {Elmore}, {Lites}, {Keller}, \&
  {Rimmele}}]{2004ApOpt..43.3817J}
{Judge}, P.~G., {Elmore}, D.~F., {Lites}, B.~W., {Keller}, C.~U., \& {Rimmele},
  T. 2004, \ao, 43, 3817

\bibitem[{{Keller} \& {Solis Team}(2001)}]{2001ASPC..236...16K}
{Keller}, C.~U. \& {Solis Team}. 2001, in Astronomical Society of the Pacific
  Conference Series, Vol. 236, Advanced Solar Polarimetry -- Theory,
  Observation, and Instrumentation, ed. M.~{Sigwarth}, 16

\bibitem[{{Kucera} {et~al.}(2015){Kucera}, {Tomczyk}, {Rybak}, {Sewell},
  {Gomory}, {Schwartz}, {Ambroz}, \& {Kozak}}]{2015IAUGA..2246687K}
{Kucera}, A., {Tomczyk}, S., {Rybak}, J., {et~al.} 2015, in IAU General
  Assembly, Vol.~29, 2246687

\bibitem[{{Kuridze} {et~al.}(2018){Kuridze}, {Henriques}, {Mathioudakis},
  {Rouppe van der Voort}, {de la Cruz Rodr{\'\i}guez}, \&
  {Carlsson}}]{2018ApJ...860...10K}
{Kuridze}, D., {Henriques}, V.~M.~J., {Mathioudakis}, M., {et~al.} 2018, \apj,
  860, 10

\bibitem[{{Ku{\v c}era} {et~al.}(2010){Ku{\v c}era}, {Ambr{\'o}z},
  {G{\"o}m{\"o}ry}, {Kozak}, \& {Ryb{\'a}k}}]{Kucera2010}
{Ku{\v c}era}, A., {Ambr{\'o}z}, J., {G{\"o}m{\"o}ry}, P., {Kozak}, M., \&
  {Ryb{\'a}k}, J. 2010, Contributions of the Astronomical Observatory
  Skalnat{\'e} Pleso, 40

\bibitem[{{Leenaarts} {et~al.}(2018){Leenaarts}, {de la Cruz Rodr{\'\i}guez},
  {Danilovic}, {Scharmer}, \& {Carlsson}}]{2018A&A...612A..28L}
{Leenaarts}, J., {de la Cruz Rodr{\'\i}guez}, J., {Danilovic}, S., {Scharmer},
  G., \& {Carlsson}, M. 2018, \aap, 612, A28

\bibitem[{{Libbrecht} {et~al.}(2019){Libbrecht}, {de la Cruz Rodr{\'\i}guez},
  {Danilovic}, {Leenaarts}, \& {Pazira}}]{2019A&A...621A..35L}
{Libbrecht}, T., {de la Cruz Rodr{\'\i}guez}, J., {Danilovic}, S., {Leenaarts},
  J., \& {Pazira}, H. 2019, \aap, 621, A35

\bibitem[{{Lites}(1987)}]{1987ApOpt..26.3838L}
{Lites}, B.~W. 1987, \ao, 26, 3838

\bibitem[{{Lites}(1991)}]{1991sopo.work..166L}
{Lites}, B.~W. 1991, in Solar Polarimetry, ed. L.~J. {November}, 166--172

\bibitem[{{Lites} \& {Ichimoto}(2013)}]{2013SoPh..283..601L}
{Lites}, B.~W. \& {Ichimoto}, K. 2013, \solphys, 283, 601

\bibitem[{{L{\"o}fdahl} {et~al.}(2018){L{\"o}fdahl}, {Hillberg}, {de la Cruz
  Rodriguez}, {Vissers}, {Scharmer}, {Hagfors Haugan}, \&
  {Fredvik}}]{2018arXiv180403030L}
{L{\"o}fdahl}, M.~G., {Hillberg}, T., {de la Cruz Rodriguez}, J., {et~al.}
  2018, arXiv e-prints, arXiv:1804.03030

\bibitem[{{Ma} {et~al.}(2008){Ma}, {Wang}, {Denker}, \&
  {Wang}}]{2008ChJAA...8..349M}
{Ma}, J., {Wang}, J.-S., {Denker}, C., \& {Wang}, H.-M. 2008, \cjaa, 8, 349

\bibitem[{McKay {et~al.}(1979)McKay, Beckman, \& Conover}]{10.2307/1268522}
McKay, M.~D., Beckman, R.~J., \& Conover, W.~J. 1979, Technometrics, 21, 239

\bibitem[{{Morosin} {et~al.}(2020){Morosin}, {de la Cruz Rodriguez}, {Vissers},
  \& {Yadav}}]{2020arXiv200614487M}
{Morosin}, R., {de la Cruz Rodriguez}, J., {Vissers}, G.~J.~M., \& {Yadav}, R.
  2020, arXiv e-prints, arXiv:2006.14487

\bibitem[{Nelder \& Mead(1965)}]{10.1093/comjnl/7.4.308}
Nelder, J.~A. \& Mead, R. 1965, The Computer Journal, 7, 308

\bibitem[{{Perez} \& {Granger}(2007)}]{2007CSE.....9c..21P}
{Perez}, F. \& {Granger}, B.~E. 2007, Computing in Science and Engineering, 9,
  21

\bibitem[{{Pietrow} {et~al.}(2020){Pietrow}, {Kiselman}, {de la Cruz
  Rodr{\'\i}guez}, {D{\'\i}az Baso}, {Pastor Yabar}, \&
  {Yadav}}]{2020arXiv200614486P}
{Pietrow}, A.~G.~M., {Kiselman}, D., {de la Cruz Rodr{\'\i}guez}, J., {et~al.}
  2020, arXiv e-prints, arXiv:2006.14486

\bibitem[{{Rodenhuis} {et~al.}(2014){Rodenhuis}, {Snik}, {van Harten},
  {Hoeijmakers}, \& {Keller}}]{2014SPIE.9099E..0LR}
{Rodenhuis}, M., {Snik}, F., {van Harten}, G., {Hoeijmakers}, J., \& {Keller},
  C.~U. 2014, in Society of Photo-Optical Instrumentation Engineers (SPIE)
  Conference Series, Vol. 9099, Polarization: Measurement, Analysis, and Remote
  Sensing XI, ed. D.~B. {Chenault} \& D.~H. {Goldstein}, 90990L

\bibitem[{{Rojo} \& {Harrington}(2006)}]{2006ApJ...649..553R}
{Rojo}, P.~M. \& {Harrington}, J. 2006, \apj, 649, 553

\bibitem[{{Samoylov} {et~al.}(2004){Samoylov}, {Samoylov}, {Vidmachenko}, \&
  {Perekhod}}]{2004JQSRT..88..319S}
{Samoylov}, A.~V., {Samoylov}, V.~S., {Vidmachenko}, A.~P., \& {Perekhod},
  A.~V. 2004, \jqsrt, 88, 319

\bibitem[{{Scharmer}(2006)}]{2006A&A...447.1111S}
{Scharmer}, G.~B. 2006, \aap, 447, 1111

\bibitem[{Scharmer {et~al.}(2003)Scharmer, Bjelksjo, Korhonen, Lindberg, \&
  Petterson}]{2003SPIE.4853..341S}
Scharmer, G.~B., Bjelksjo, K., Korhonen, T.~K., Lindberg, B., \& Petterson, B.
  2003, in Innovative Telescopes and Instrumentation for Solar Astrophysics,
  ed. S.~L. Keil \& S.~V. Avakyan, Vol. 4853, International Society for Optics
  and Photonics (SPIE), 341 -- 350

\bibitem[{{Scharmer} {et~al.}(2019){Scharmer}, {L{\"o}fdahl}, {Sliepen}, \& {de
  la Cruz Rodr{\'\i}guez}}]{2019A&A...626A..55S}
{Scharmer}, G.~B., {L{\"o}fdahl}, M.~G., {Sliepen}, G., \& {de la Cruz
  Rodr{\'\i}guez}, J. 2019, \aap, 626, A55

\bibitem[{{Scharmer} {et~al.}(2008){Scharmer}, {Narayan}, {Hillberg}, {de la
  Cruz Rodriguez}, {L{\"o}fdahl}, {Kiselman}, {S{\"u}tterlin}, {van Noort}, \&
  {Lagg}}]{2008ApJ...689L..69S}
{Scharmer}, G.~B., {Narayan}, G., {Hillberg}, T., {et~al.} 2008, \apjl, 689,
  L69

\bibitem[{{Selbing}(2005)}]{Selbing2005}
{Selbing}, J. 2005, Master's thesis, Stockholm Univ., arXiv:1010.4142

\bibitem[{{Semel}(2003)}]{2003A&A...401....1S}
{Semel}, M. 2003, \aap, 401, 1

\bibitem[{{Serkowski}(1974)}]{1974MExP...12..361S}
{Serkowski}, K. 1974, Methods of Experimental Physics, 12, 361

\bibitem[{{Snik} {et~al.}(2015){Snik}, {van Harten}, {Alenin}, {Vaughn}, \&
  {Tyo}}]{2015SPIE.9613E..0GS}
{Snik}, F., {van Harten}, G., {Alenin}, A.~S., {Vaughn}, I.~J., \& {Tyo}, J.~S.
  2015, in Society of Photo-Optical Instrumentation Engineers (SPIE) Conference
  Series, Vol. 9613, Polarization Science and Remote Sensing VII, 96130G

\bibitem[{{Snik} {et~al.}(2012){Snik}, {van Harten}, {Navarro}, {Groot},
  {Kaper}, \& {de Wijn}}]{2012SPIE.8446E..25S}
{Snik}, F., {van Harten}, G., {Navarro}, R., {et~al.} 2012, in Society of
  Photo-Optical Instrumentation Engineers (SPIE) Conference Series, Vol. 8446,
  Ground-based and Airborne Instrumentation for Astronomy IV, 844625

\bibitem[{{Tomczyk} {et~al.}(2010){Tomczyk}, {Casini}, {de Wijn}, \&
  {Nelson}}]{2010ApOpt..49.3580T}
{Tomczyk}, S., {Casini}, R., {de Wijn}, A.~G., \& {Nelson}, P.~G. 2010, \ao,
  49, 3580

\bibitem[{{van der Walt} {et~al.}(2011){van der Walt}, {Colbert}, \&
  {Varoquaux}}]{2011CSE....13b..22V}
{van der Walt}, S., {Colbert}, S.~C., \& {Varoquaux}, G. 2011, Computing in
  Science and Engineering, 13, 22

\bibitem[{{van Noort} \& {Rouppe van der Voort}(2008)}]{2008A&A...489..429V}
{van Noort}, M.~J. \& {Rouppe van der Voort}, L.~H.~M. 2008, \aap, 489, 429

\bibitem[{{Vissers} {et~al.}(2020){Vissers}, {Danilovic}, {de la Cruz
  Rodriguez}, {Leenaarts}, {Morosin}, {Diaz Baso}, {Reid}, {Pomoell}, {Price},
  \& {Inoue}}]{2020arXiv200901537V}
{Vissers}, G.~J.~M., {Danilovic}, S., {de la Cruz Rodriguez}, J., {et~al.}
  2020, arXiv e-prints, arXiv:2009.01537

\bibitem[{{Vissers} {et~al.}(2019){Vissers}, {de la Cruz Rodr{\'\i}guez},
  {Libbrecht}, {Rouppe van der Voort}, {Scharmer}, \&
  {Carlsson}}]{2019A&A...627A.101V}
{Vissers}, G.~J.~M., {de la Cruz Rodr{\'\i}guez}, J., {Libbrecht}, T., {et~al.}
  2019, \aap, 627, A101

\end{thebibliography}
\end{document}